\documentclass[twocolumn]{aastex63}
\usepackage{graphics,epsf}
\usepackage{amsmath}                
\usepackage{amsfonts}               
\usepackage{amssymb}                
\usepackage{epsfig}                 
\usepackage{appendix}
\usepackage{graphicx}
\usepackage{float}
\usepackage{color}
\usepackage{multirow}
\usepackage{colortbl}
\usepackage[para,online,flushleft]{threeparttable}

\newcommand{\km}{{~\rm km}}
\newcommand{\s}{{~\rm s}}

\newcommand{\erg}{{~\rm erg}}
\newcommand{\yr}{{~\rm yr}}

\begin{document}

\title{Spin-orbit misalignment from triple-star common envelope evolution}

\email{soker@physics.technion.ac.il}

\author{Noam Soker}
\affiliation{Department of Physics, Technion – Israel Institute of Technology, Haifa 3200003, Israel}
\affiliation{Guangdong Technion Israel Institute of Technology, Guangdong Province, Shantou 515069, China}

\begin{abstract}
I study a triple star common envelope evolution (CEE) of a tight binary system that is spiraling-in inside a giant envelope and launches jets that spin-up the envelope with an angular momentum component perpendicular to the orbital angular momentum of the triple star system. This occurs when the orbital plane of the tight binary system and that of the triple star system are inclined to each other, so the jets are not along the triple star orbital angular momentum. The merger of the tight binary stars also tilts the envelope spin direction. If the giant is a red supergiant (RSG) star that later collapses to form a black hole (BH) the BH final spin is misaligned with the orbital angular momentum. Therefore, CEE of neutron star (NS) or BH tight binaries with each other or with one main sequence star (MSS) inside the envelope of an RSG, where the jets power a common envelope jets supernova (CEJSN) event, might end with a NS/BH–NS/BH close binary system with spin-orbit misalignment. Such binaries can later merge to be gravitational waves sources.  I list five triple star scenarios that might lead to spin-orbit misalignments of NS/BH–NS/BH binary systems, two of which predict that the two spins be parallel to each other. In the case of a tight binary system of two MSSs inside an asymptotic giant branch star the outcome is an additional non-spherical component to the mass loss with the formation of a `messy’ planetary nebula. 
\end{abstract}


\section{Introduction} 
\label{sec:intro}

Common envelope jets supernovae (CEJSNe) are binary and multiple stellar systems where during a common envelope evolution (CEE) a neutron star (NS) or a black hole (BH), here after NS/BH, spiral-in inside the envelope of a red supergiant (RSG) star, accrete mass via an accretion disk, and launch energetic jets (e.g., \citealt{Gilkisetal2019, SokeretalGG2019, GrichenerSoker2019a, Schroderetal2020, GrichenerSoker2021}). The NS/BH then enters the core and launches very energetic jets. Strictly speaking, if the NS/BH does not enter the core it is a CEJSN impostor. However, I will here use the term CEJSN to refer also to CEJSN impostors.  

The very small radius of the NS/BH allows for three processes that are essential to form CEJSNe that are powerful as core collapse supernovae (CCSNe) are and even more. (1) Because of the radial density gradient in the RSG envelope the mass that the NS/BH accretes has a specific angular momentum that is sufficiently high to form an accretion disk around the compact NS/BH (e.g.,  \citealt{ArmitageLivio2000, Papishetal2015, SokerGilkis2018,  LopezCamaraetal2019, LopezCamaraetal2020MN}; for more on the question of accretion disk formation and on mass accretion rate see \citealt{Hilleletal2021} and references therein). The small radius of the NS/BH ensures the formation of an accretion disk. 
(2) The small radius implies a deep potential well that results in very fast jets $v_{\rm j} > 0.1 c$, namely, can be highly relativistic in particular with a BH as the  mass-accreting object.
These jets might host processes as those that take place in jets in CCSNe and in NS binary merger, e.g., production of energetic neutrinos \citep{GrichenerSoker2021} and r-process nucleosynthesis (e.g., \citealt{GrichenerSoker2019a}). 
(3) The small radius of the NS/BH allows for an efficient neutrino-cooling of the in-flowing mass that in turns allows for a very high mass accretion rate $\dot M_{\rm acc}$, i.e., above the Eddington mass accretion rate (\citealt{HouckChevalier1991, Chevalier1993, Chevalier2012}; the condition for efficient neutrino cooling $\dot M_{\rm acc} \ga 10^{-3} M_\odot \yr^{-1}$ holds in CEJSNe).
A BH can as well accrete the mass with a large portion of its energy (e.g., \citealt{Pophametal1999}).
 
The NS/BH that spirals-in inside the RSG envelope might either eject the envelope before it reaches the core or it might enter the core (e.g., \citealt{SokeretalGG2019}). If the NS/BH  ejects the entire envelope before entering the core it leaves the core to later explode as a second (or third in some triple-star systems) CCSN in the system. This evolutionary route leaves a NS/BH - NS/BH system, bound or unbound. In the other type of evolution the NS/BH continues with the CEE into the RSG core and destroys the core to form a massive accretion disk that launches very energetic jets (e.g., \citealt{GrichenerSoker2019a}). This might end with an explosion energy of up to $\simeq {\rm several} \times 10^{52} \erg$ that powers a bright light curve that lasts for up to few years and where the ejecta collides with a very massive circumstellar matter of several solar masses and more. Due to the long lasting and energetic light curve with several peaks this explosion would be classified as a peculiar supernova (SN; e.g., \citealt{SokeretalGG2019, Schroderetal2020}). 
The enigmatic SN~iPTF14hls \citep{Arcavietal2017} and SN~2020faa \citep{Yangetal2021} might be CEJSN events \citep{SokerGilkis2018}. A different type of CEJSN event \citep{SokeretalGG2019} might account for the fast-rising blue optical transient AT2018cow (e.g., \citealt{Prenticeetal2018, Marguttietal2019}).
I also note that the core-NS/BH merger might be a source of gravitational waves (e.g., \citealt{Ginatetal2020}). 

The interaction of the jets with the RSG envelope unbinds a large fraction of the RSG envelope to the degree that the energetic jets  can supply an amount of energy in addition to the orbital energy of the spiralling-in NS/BH and make the CEE efficiency parameter larger than unity $\alpha_{\rm CE} > 1$. Some scenarios require indeed values of $\alpha_{\rm CE} > 1$ (e.g. \citealt{Fragosetal2019, BroekgaardenBerger2021,  Garciaetal2021, Zevinetal2021}).

The large fraction of triple and higher order star systems among massive stars (e.g., \citealt{Sanaetal2014, MoeDiStefano2017}) and the rich triple star interaction outcomes (e.g., \citealt{Toonenetal2021}) have motivated studies of CEJSNe in triple stars systems (e.g., \citealt{Soker2021Double, Soker2021NSNSmerger, AkashiSoker2021}) as well as other triple-star CEE (e.g., \citealt{SabachSoker2015, Hilleletal2017, Schreieretal2019inclined, ComerfordIzzard2020, GlanzPerets2021, SokerBear2021Parasite}). 

In section \ref{subsec:DepositionJets} that contains the main calculation of this study I describe another type of outcome from the interaction of jets with the RSG envelope in triple-star CEE. Although I concentrate on NS/BH spiralling-in inside RSG envelopes, some of the processes are relevant to main sequence stars (MSSs) spiralling-in inside the envelope of asymptotic giant branch (AGB) stars. In the main process I study a tight (inner) binary system has its orbital plane inclined to that of the triple system (i.e., the orbital plane of the giant and the tight binary system). After the tight binary enters the giant envelope it accretes mass and launches jets. Because the orbital plane of the tight binary system and the orbital plane of the triple star system are inclined to each other, the jets are inclined to the orbital angular momentum axis of the triple system. 
This general process holds for a tight binary system of two MSSs where one or two MSSs launch the jets, and for tight binary systems where one or the two stars are NS/BHs, therefore a NS/BH launches the jets. The result is that the jets impart angular momentum to the giant envelope that is perpendicular to the orbital angular momentum of the triple star system.

I describe three triple-star scenarios in section \ref{sec:JetsMisalignment} that might lead to spin-orbit misalignment in post-CEE binary systems. 
In section \ref{sec:Others} I describe two other processes by which triple-star CEE might lead to spin-orbit misalignment in post-CEE systems.
I summarise this study in section \ref{sec:Summary} with the main results and a short discussion on the implications to NS/BH binary merger that are sources of gravitational waves. 

\section{Misaligned angular momentum by jets} 
\label{sec:JetsMisalignment}

\subsection{The triple CEE settings} 
\label{subsec:Settings}

\begin{figure}[t!]
\includegraphics[trim=5.5cm 5.3cm 0.0cm 2.0cm ,clip, scale=0.55]{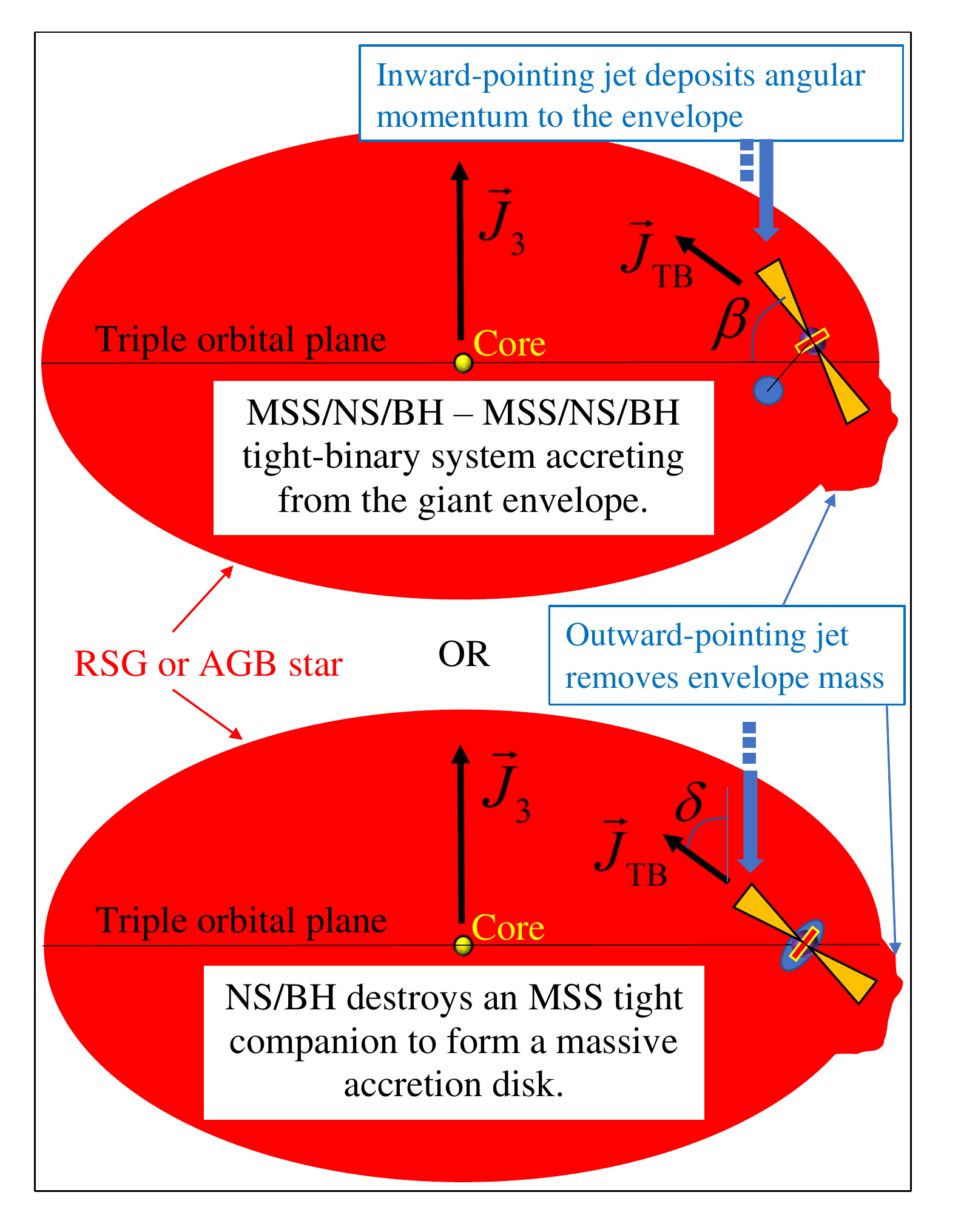}
\caption{A schematic diagram of the inclined jets that a tight binary system launches as it spirals-in inside the envelope of an RSG, or possibly through the envelope of an AGB star when the tight binary system is of two MSSs. In the upper panel accretion from the giant envelope forms the accretion disk around one of the tight binary stars or a disk around each of the two tight binary stars. In the lower panel the more compact star in the tight binary system is a NS/BH that destroys the MSS to form an accretion disk that launches the jets. The vectors $\vec{J}_3$ and $\vec{J}_{\rm TB}$ represents the orbital angular momentum of the triple-star system and of the tight binary system, respectively. The angular momentum that the jets deposit to the giant envelope points out of the figure. The panels present the orbital phase $\phi=0$. 
Abbreviation: AGB: asymptotic giant branch; BH: black hole; MSS: main sequence star; NS: neutron star; RSG: red super giant. }
 \label{fig:SchematicScenario}
 \end{figure}

I deal with a tight binary system that enters the giant envelope and the orbital plane of the tight binary system is inclined to the triple-star orbital plane. The tight binary system can be NS/BH-NS/BH binary, NS/BH-MSS binary, or MSS-MSS binary as I present schematically in Fig. \ref{fig:SchematicScenario}. 
In the first setting (upper panel of Fig. \ref{fig:SchematicScenario}) the tight binary system accretes mass from the inhomogeneous medium of the RSG envelope (for simulations of a binary system accreting from an ambient medium, including inclined orbits, see, e.g., \citealt{Comerfordetal2019}). As a result of that an accretion disk forms around one or each of the two tight binary components and the accretion disk (or two disks) launches jets. Both the orbital angular momentum of the triple-star system $\vec{J}_3$ and that of the tight binary system $\vec{J}_{\rm TB}$ determine the direction of the angular momentum of the accretion disks. I expect the angular momentum of the accretion disk, which is the direction of the jets' axis that it launches, to be in between the two.

In the second setting the more compact component of the tight binary system, a NS/BH (or very rarely the more compact MSS in the case of two MSSs), merges with an MSS companion and destroys it to form an accretion disk around the NS/BH. In that case the accretion disk launches the jets in the same direction as  $\vec{J}_{\rm TB}$. I describe this scenario in section \ref{subsec:DoubleCEJSN}.

\subsection{Angular momentum from jets} 
\label{subsec:DepositionJets}

I here crudely calculate the change in the envelope angular velocity because of the interaction of the jets with the envelope as I draw schematically in the upper panel of Fig. \ref{fig:SchematicScenario}. 
To facilitate such a calculation I make the following assumptions.
\newline
(1) The jets are substantially inclined to both the orbital plane and to $\vec{J}_3$, i.e., $ 15^\circ \la \beta \la 75^\circ$. 
\newline
(2) The jets operate in a negative jet feedback mechanism (e.g., \citealt{Soker2016Rev, Gricheneretal2021, Hilleletal2021}). 
\newline
(3) As a consequence of assumption 2 above, the jets are strong enough to unbind the envelope outside the orbit of the tight binary system, but not for the jet that points inward to break out from the envelope. For the asymmetric influence of the two jets on the envelope I simply take a factor of 
\begin{equation}
F_{\rm asymm} =  \cos \beta. 
\label{eq:Fasymm}
\end{equation}
This vanishes for jets that are aligned with $\vec{J}_3$.
\newline
(4) I assume that convection in the giant envelope rapidly distributes the angular momentum in the envelope inner to the tight binary orbit around the giant core, such that the inner envelope more or less reaches solid body rotation. 
\newline
(5) I take the triple-orbital phases $\phi=0$ and $\phi =\pi$ to represent the times when one jet points directly inward and one jet points directly outward (Fig. \ref{fig:SchematicScenario} is at $\phi=0$). When the orbital phase is such that the two opposite jets are more or less symmetric with respect to the core the two jets deposit opposite angular momentum around the center of the giant and the net angular momentum deposition vanishes. This occurs near phases $\phi =\pi/2$ and $\phi =3 \pi/2$.
For the factor that represents this variation along the orbit I take $F_{\rm orbit,\phi} =  \vert \cos \phi \vert$, which over an orbit introduces an average factor of 
\begin{equation}
F_{\rm orbit}=\frac{2}{\pi}.
\label{eq:orbita}
\end{equation}
I take an absolute value $\vert \cos \phi \vert$  because on opposite sides of the core the two jets (one above and one below the equatorial plane) exchange the roles of unbinding envelope gas and depositing angular momentum to the envelope. 

I assume that at triple orbital phases $\phi=0$ and $\phi =\pi$ the outward-pointing jet removes mass from the envelope and therefore it does not contribute angular momentum to the envelope that stays bound at that time. I take the momentum that the outward-pointing jet imparts to the ejected envelope mass to be 
\begin{equation}
dp_{\rm ej} \approx v_{\rm esc} dM_{\rm env},  
\label{eq:dPej}
\end{equation}
where $v_{\rm esc}$ is the escape velocity from the giant envelope and $dM_{\rm env}$ is the mass that the jet removes from the envelope. 
If the inward pointing jet deposits a similar momentum to the envelope, the angular momentum it deposits around the core of the giant is $\simeq a \sin \beta dp_{\rm ej}$. With the averaging over an orbit (equation \ref{eq:orbita}) and the asymmetrical factor (equation \ref{eq:Fasymm}), the average (over an orbit) angular momentum that the jets deposit to the envelope as they remove mass $dM_{\rm env}$ is 
\begin{equation}
d J_{\rm 2j} \approx \frac{2}{\pi} \cos \beta \sin \beta a v_{\rm esc} dM_{\rm env} .  
\label{eq:dJ2j}
\end{equation}
The direction of the angular momentum $d J_{\rm 2j}$ is perpendicular to both $\vec{J}_3$ and $\vec{J}_{\rm TB}$, and it points out of Fig. \ref{fig:SchematicScenario}.

For the moment of inertia of the envelope I take  
\begin{equation}
I_{\rm env} = \eta M_{\rm env} R^2 \simeq 0.2 M_{\rm env} R^2.
\label{eq:Ienv}
\end{equation}
As examples, for $\rho \propto r^{-2}$ and  $\rho \propto r^{-2.5}$ one finds $\eta=2/9$ and $2/15$, respectively. 
 The additional angular velocity due to the jets is then 
\begin{equation}
d \Omega_{\rm 2j} \simeq \frac{d J_{\rm 2j}}{I_{\rm env}}
\approx 
\frac{\sin 2 \beta}{\pi \eta} \frac{v_{\rm esc}}{a} 
\frac {dM_{\rm env}}{M_{\rm env}},  
\label{eq:dOmega}
\end{equation}
where I substituted $R \simeq a$ under assumption (4) above. 
The escape velocity is $v_{\rm esc} = [2 G (M_{\rm core}+M_{\rm env} + M_{\rm TB})/a]^{1/2}$, where $M_{\rm core}$ and $M_{\rm TB}$ are the core and the tight binary masses, respectively. Because I assume a massive core and a massive tight binary system, to the present order of magnitude derivation I consider the escape velocity to vary as 
$v_{\rm esc} = [2 G (M_{\rm core}+M_{\rm env,f} + M_{\rm TB})/a]^{1/2}$, namely, I neglect the reduction in the envelope mass in the expression for the escape velocity and take the envelope mass to be its final mass $M_{\rm env,f}$ at some late time.

To proceed I take the envelope density to be $\rho=\rho_0(r/R_0)^{-2}$, where zero denotes at the initial giant envelope surface. This profile is a fare approximation to an extended giant envelope. Integration gives for the envelope mass inner to radius $r$, $M_{\rm env}(r)=(r/R_0)M_{\rm env,0}$. With the assumption that the co-rotating envelope extends from the base of the envelope to $r\simeq a$, the relevant part of the envelope is inner to the orbit of the tight binary that is given by $a=R_0(M_{\rm env}/M_{\rm env,0})$.
I now integrate the two sides of equation (\ref{eq:dOmega}) with the above expression for $a$
\begin{eqnarray}
\begin{aligned}
\int_0^{\Omega_{\rm 2j,f}}  d \Omega_{\rm 2j} 
& \approx 
- \frac{\sin 2 \beta}{\pi \eta}  
\left[2G \left( M_{\rm core} + M_{\rm env,f} + M_{\rm TB}) \right) \right]^{1/2} 
\\ & \times 
R^{-3/2}_0
\int^{M_{\rm env,f}}_{M_{\rm env,0}}
\left( \frac {M_{\rm env}}{M_{\rm env,0}} \right)^{-5/2}  
\frac {dM_{\rm env}}{M_{\rm env,0}} ,
\end{aligned}
\label{eq:OmegaF1}
\end{eqnarray}
where the minus sign in the right hand side comes from the definition of the removed envelope mass as a positive quantity. Taking $\eta=2/9$ for the assumed density profile $\rho \propto r^{-2}$, the integration of equation (\ref{eq:OmegaF1}) yields  
\begin{equation}
\Omega_{\rm 2j,f } 
\approx 1.3 \Omega_{\rm crit} \sin 2 \beta 
\left[ \left( \frac {M_{\rm env,f}}{M_{\rm env,0}} \right)^{-3/2} -1 \right] 
\left( \frac {R_{\rm f}}{R_0} \right)^{3/2},  
\label{eq:OmegaF2}
\end{equation}
where 
\begin{equation}
\Omega_{\rm crit} = \sqrt {\frac {G (M_{\rm core}+ M_{\rm env,f} + M_{\rm TB})}{R^3_{\rm f}}}, 
\label{eq:OmegaCrit}
\end{equation}
is the critical angular velocity of the envelope at a final envelope radius $R_{\rm f}$.   
Taking $M_{\rm env,f} = M_{\rm env,0} (R_{\rm f}/R_0) \ll M_{\rm env,0}$ yields the approximate final expression for the ratio of angular momentum that the jets deposit to the envelope to the critical angular velocity of the envelope 
\begin{equation}
\frac {\Omega_{\rm 2j,f } } { \Omega_{\rm crit} }
\approx \sin 2 \beta .
\label{eq:OmegaF3}
\end{equation}

Although the derivation is very crude, it nonetheless shows that the jets can cause the final envelope to have a non negligible angular momentum component that is perpendicular to both the orbital angular momentum of the triple star system and to that of the tight binary, $\vec{J}_3$ and $\vec{J}_{\rm TB}$, respectively. 

\begin{figure*}[t!]
\includegraphics[trim=1.5cm 8.3cm 0.0cm 2.0cm ,clip, scale=0.95]{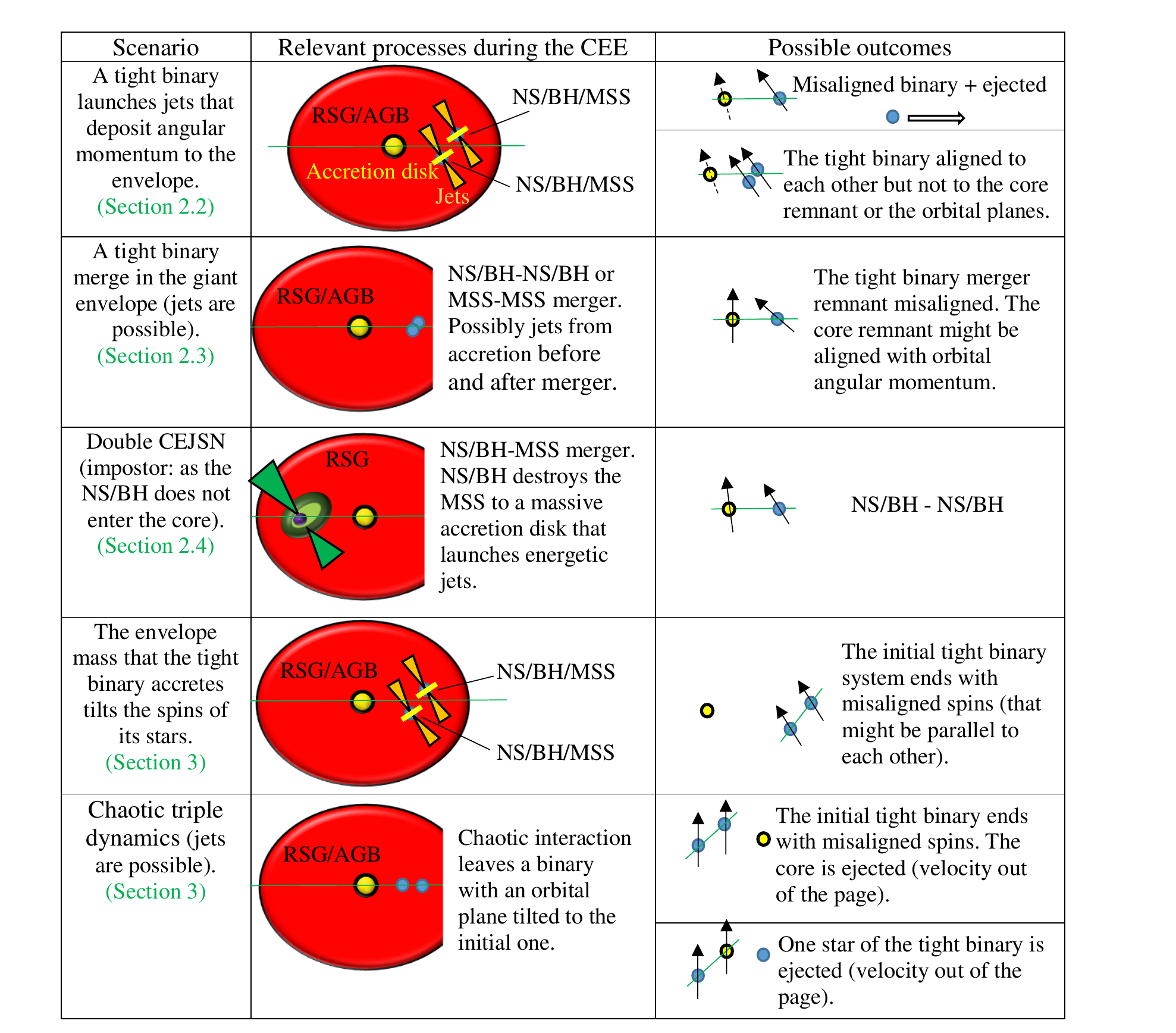}
\caption{ A list of the scenarios for the formation of spin-orbit misalignment in post-CEE binary and triple systems that I study here. 
Dashed arrows indicate spins that are not in the plane of the page. The initial triple star orbital plane is a plane perpendicular to the page and horizontal as the grin lines in the middle column show. The green line in each panel of the right column represents the orbital plane of the final binary system that has the misaligned spin(s). 
A yellow-black circle is for the core and its remnant. 
Abbreviation: AGB: asymptotic giant branch; BH: black hole; MSS: main sequence star; NS: neutron star; RSG: red super giant.
}
 \label{fig:cases}
 \end{figure*}

I summarise this scenario in the first row (below the titles row) of Fig. \ref{fig:cases}. 
 
This process where the jets cause spin-orbit misalignment is speculative. Three-dimensional hydrodynamical simulations should find whether it can work or not. There is a positive hint from the inclined-jets simulation of \cite{Schreieretal2019inclined}, although this simulation is  for jets that a tight binary system of low-mass main sequence stars launches. The jets initial velocity was only $500 \km \s^{-1}$. Nonetheless, their results demonstrate the asymmetry of outflows that the two jets induce in the envelope. The upper panel of figure 5 of \cite{Schreieretal2019inclined} show that the flow inside the envelope that the jet that expands against the orbital motion of the tight binary system induces a larger flow in the common envelope than the flow that the jet that expands in the same direction as that of the orbital motion induces. The present process is built on a much large asymmetrical jet-envelope interaction as one jet removes mass from the envelope. This takes place with more energetic jets than the jets that \cite{Schreieretal2019inclined}  simulated.   

\subsection{Angular momentum from the tight binary merger} 
\label{subsec:DepositionTB}
 
The merger of the binary system inside the envelope implies that most of the initial binary angular momentum ends in the envelope. 
I take the two components of the tight binary system to have initial masses (when entering the envelope) of $M_2$ and $M_3$ and an initial circular orbit of radius $a_{\rm TB, 0}$. The contraction of the tight binary systems occurs as it spirals in. I assume that the merger process deposits most of the initial angular momentum of the tight binary system inner to some radius $a_{\rm mer}$. Using the same expression as before with an envelope density profile of $\rho \propto r^{-2}$, I find the contribution of the merger at $a_{\rm mer}$ to the angular velocity component perpendicular to the triple-star angular momentum (perpendicular to $\vec{J}_3$) to be 
\begin{eqnarray}
\begin{aligned}
\frac{\Omega_{\rm mer}}{\Omega_{\rm crit} }
& \simeq \frac{9}{2}
\frac{M_2 M_3}
{\left[ \left( M_{\rm core} + M_{\rm env,f} + M_{\rm TB} \right)
M_{\rm TB}
\right]^{1/2} }
\\ & \times
\frac {1}{M_{\rm env} (a_{\rm mer})} \left( \frac{a_{\rm TB,0}}{a_{\rm mer}} \right)^{1/2}  \sin \delta,
\end{aligned}
\label{eq:OmegaTB}
\end{eqnarray}
where $\delta$ is the angle between $\vec{J}_3$ and $\vec{J}_{\rm TB}$. This contribution to the misalignment between the merger remnant angular momentum and the orbital angular momentum can be non-negligible. I summarise this scenario in the second row (below titles) of Fig. \ref{fig:cases}.  

In all CEE scenarios that I study here the largest contribution to the envelope rotation is likely to come from the triple-star orbital angular momentum. However, the other effects might be non-negligible and substantially tilt the final spin of the RSG remnant.  
I conclude that the final envelope that stays bound might have its angular momentum misaligned with that of the triple system $\vec{J}_3$, i.e., with the final orbital angular momentum.
If now the giant collapses to form a BH, some of the envelope gas might be accreted onto the newly born BH, and the BH ends with an angular momentum axis that is misaligned to the initial triple-star angular momentum, which at the end is the orbital angular momentum of the newly born BH with either the tight binary system or with its merger product. A quantitative derivation of the final BH angular momentum requires detailed numerical simulations with a large dynamical range.  

\subsection{Angular momentum from double CEJSN} 
\label{subsec:DoubleCEJSN}

In a double CEJSN \citep{Soker2021Double} the tight binary system is a NS/BH and a MSS that merge inside the envelope. The NS/BH destroys the MSS to form a massive accretion disk that launches very energetic jets (lower panel of  Fig. \ref{fig:SchematicScenario}). Later, the NS/BH remnant might enter the core of the RSG or else remove the entire envelope so that the core explodes as a CCSN and leaves a NS/BH remnant. (Strictly speaking, if the NS/BH does not enter the core it is a double CEJSN impostor.) I here consider the case where the orbital plane of the tight binary system and the triple system are not parallel to each other, and the core does explode to leave a NS/BH bound to the older NS/BH. I summarise this scenario in the third scenario in Fig. \ref{fig:cases}. 

The spin of the older NS/BH will be inclined to the orbital angular momentum of the final NS/BH-NS-BH binary. The spin of the newly born NS/BH might be parallel to the orbital angular momentum or inclined by the same mechanism of jets-deposition of angular momentum to the leftover envelope that I derived in section \ref{subsec:DepositionJets} (neglecting natal kick velocity outside the orbital plane of the newly born NS/BH).

\section{Other cases of misalignment by triple CEE} 
\label{sec:Others}

In the three scenarios that I described in sections \ref{subsec:DepositionJets} - \ref{subsec:DoubleCEJSN} the binary remnant that has one or two of its spins misaligned with the orbital angular momentum is composed of the remnant of the RSG/AGB core and one or two of the components of the tight binary system or their merger remnant. 
I describe here two scenarios that might leave the initial tight binary system as the final binary with spin-orbit misalignment. 

The two components of the inclined tight binary system accrete mass from the RSG/AGB envelope that has its angular momentum in the same direction as that of the triple star orbital angular momentum. This angular momentum is inclined to the orbital angular momentum of the tight binary system, and might lead to spin-orbit misalignment in the tight binary system. If the two components of the tight binary system are similar, e.g., two similar NSs or two similar BHs, at the final state they might have their spins parallel to each other, although misaligned with the tight binary orbital angular momentum.
I describe this scenario in the forth row of Fig. \ref{fig:cases}.
 
Another scenario (fifth in Fig. \ref{fig:cases}) involves a chaotic triple dynamics of the core with the tight binary system. This chaotic interaction might eject the lowest-mass star of the three and leaves the two others in a bound state (e.g., \citealt{GlanzPerets2021}). Because of the chaotic interaction any small initial misalignment might lead to a final binary with its orbital angular momentum inclined to the initial angular momentum of the triple star system. 
If the initial orbital planes of the tight binary system and that of the triple stellar system are (almost) parallel to each other, most likely the two spins will be parallel to each other, although misaligned with the orbital angular momentum.

\section{Discussion and summary} 
\label{sec:Summary}

The main new result of this study is the demonstration that jets that a tilted tight binary system launches in a CEE can spin-up the envelope and add an angular velocity component that is perpendicular to both the initial triple star system orbital angular momentum $\vec{J}_3$ and to the tight binary orbital angular momentum $\vec{J}_{\rm TB}$ (upper panel of Fig. \ref{fig:SchematicScenario} for the flow structure).
Under the assumptions that I listed at the beginning of section \ref{subsec:DepositionJets} I crudely calculated this angular velocity component to be a substantial fraction of the critical angular velocity of the leftover envelope (equation \ref{eq:OmegaF3}). 
A contraction of the tight binary system in the CEE also deposits angular momentum to the envelope along the $\vec{J}_{\rm TB}$ direction (lower panel of Fig. \ref{fig:SchematicScenario}). Equation (\ref{eq:OmegaTB}) gives the value of the additional angular velocity component perpendicular to $\vec{J}_3$ in case of a merger of the tight binary system.

In all CEE scenarios that I studied the largest contribution to the giant envelope rotation is from the triple-star orbital angular momentum because the tight binary system must transfer angular momentum to the envelope when it spirals-in towards the core of the RSG. This is similar to the case with binary star CEE. This largest spin-up process has its angular momentum along $\vec{J}_3$. Nonetheless, the other spin-up processes of the RSG envelope that I studied here for inclined-triple star CEE evolution might be non-negligible and substantially tilt the final spin of the RSG remnant with respect to $\vec{J}_3$.

The case of an AGB star that engulfs a tight binary system of two MSSs or of a white dwarf with a MSS is relevant to the formation of planetary nebulae, and in particular to planetary nebulae with `messy' morphologies, i.e, that lack any kind of symmetry.
`Messy' planetary nebulae most like result from triple star interaction (e.g., \citealt{BearSoker2017, Danehkaretal2018, Jonesetal2019, Miszalskietal2019, Schreieretal2019inclined, RechyGarciaetal2020, Henneyetal2021}). 
The spinning-up of the AGB envelope to have a rotation that is inclined to the triple star orbital angular momentum that I studied here adds to the `messy' mass loss morphology. 

In the same manner, the spinning-up of the RSG envelope by the jets and by the merger inside the envelope in the case of CEJSNe in triple star systems add to the highly non-spherical mass ejection by the jets. The early outflow and the later more energetic ejecta will be `messy'. The outflow geometry influences radiative transfer in the ejecta and the collision of fast parcels of gas with slow ones, both processes of which affect the light curve. I expect the light curve as a result of the clumpy and messy ejecta morphology that lack any symmetry to be non-monotonic, i.e., be bumpy.    
  
The main aim of this study is to account for spin-orbit misalignment in NS/BH-NS/BH binary systems. The five scenarios that I list in Fig. \ref{fig:cases} apply mainly to the formation of these binary systems. 
I did not include the effect of the natal kick on the newly born NS/BH (in particular for a NS) that also lead to misalignment and in any case influences it (e.g., \citealt{Gerosaetal2018, Wysockietal2018, Fragioneetal2021}). Interestingly, \cite{Fragioneetal2021} require a large CEE parameter, $\alpha_{\rm CE} > 1$, something that jets in CEE can account for (section \ref{sec:intro}).
 
I specifically note that some of the scenarios predict that the two spins be parallel to each other although they are misaligned with the orbital angular momentum (section \ref{sec:Others} and lower two rows of Fig. \ref{fig:cases}). As well, the merger scenario (section \ref{subsec:DepositionTB}) predicts that in some cases only the merger remnant of the tight binary system has its spin misaligned with the orbital angular momentum, but not the NS/BH remnant of the RSG core. 

It is possible that a large fraction and even most NS/BH-NS/BH merger sources of gravitational waves result from CEE (e.g., \citealt{Belczynskietal2020, BroekgaardenBerger2021, ShaoLi2021}; but note the uncertainties, e.g., \citealt{Belczynskietal2021}). If this is the case, then my study suggests that triple star evolution might explain many (but not all) of the NS/BH-NS/BH merging systems that have spin-orbit misalignment. 
Future population synthesis studies will have to find the fractional contribution of the different scenarios that I list in Fig. \ref{fig:cases}.

\section*{Acknowledgments}

I thank Aldana Grichener and Avishai Gilkis for helpful discussions and comments, and an anonymous referee for a useful comment. 
This research was supported by a grant from the Israel Science Foundation (769/20).

\textbf{Data availability}
The data underlying this article will be shared on reasonable request to the corresponding author.  


\label{lastpage}
\end{document}